\documentclass[preprint,12pt]{elsarticle}

\usepackage{amssymb,bm}
\usepackage{amsmath}
\usepackage{array,threeparttable,booktabs,multirow}
\usepackage{graphicx}
\usepackage[ruled]{algorithm2e}
\usepackage[noend]{algpseudocode}
\usepackage[dvipsnames]{xcolor}
\usepackage{setspace}
\usepackage[colorlinks=true, allcolors=blue]{hyperref}
\usepackage{pifont}
\usepackage{makecell}
\usepackage{diagbox}
\usepackage{setspace}
\usepackage{float}   
\usepackage{subfigure} 
\usepackage{enumitem}
\usepackage{orcidlink}
\usepackage{fancyhdr}

\journal{a journal}

\begin{document}

\thispagestyle{fancy}
\begin{frontmatter}



\title{Probabilistic Net Load Forecasting for High-Penetration RES Grids Utilizing Enhanced Conditional Diffusion Model}


\author[label1]{Yixiang Huang} %% Author name

\affiliation[label1]{organization={School of Electrical and Electronic Engineering, Huazhong University of Science and Technology},%Department and Organization
            addressline={1037 Luoyu Road}, 
            city={Wuhan},
            postcode={430074}, 
            state={Hubei},
            country={China}}

\author[label1]{Jianhua Pei} %% Author name

\author[label1]{Luocheng Chen} %% Author name

\author[label1]{Zhenchang Du} %% Author name

\author[label1]{Jinfu Chen\corref{cor1}} %% Author name
\ead{chenjinfu@hust.edu.cn}

\cortext[cor1]{Corresponding author}

\author[label2]{Zirui Peng} %% Author name

\affiliation[label2]{organization={State Grid Wuhan Electric Power Company},%Department and Organization
            addressline={1053 Jiefang Avenue}, 
            city={Wuhan},
            postcode={430000}, 
            state={Hubei},
            country={China}}

%% Abstract
\begin{abstract}
The proliferation of intermittent distributed renewable energy sources (RES) in modern power systems has fundamentally compromised the reliability and accuracy of deterministic net load forecasting. Generative models, particularly diffusion models, demonstrate exceptional potential in uncertainty quantification for scenario forecasting. Nevertheless, their probabilistic predictive capabilities and conditional bootstrapping mechanisms still remain underexplored. In this paper, a day-ahead probabilistic net load forecasting framework is developed by systematically quantifying epistemic uncertainty and aleatoric variability using the feature-informed enhanced conditional diffusion model (ECDM). The ECDM architecture implements the net load distribution generation process using an imputation-based conditional diffusion model, where multi-modal conditional inputs, such as weather and calendar data, are fused via cross-attention mechanisms. Specifically, historical net load profiles are utilized to guide the reverse diffusion trajectory through non-parametric imputation operators preserving spatial-temporal integrity. To capture periodic characteristics, a novel weekly arrangement method is also introduced, while an unconditional model is integrated to ensure diversity in the generated scenarios. Subsequently, the maximum probabilistic points and probability intervals of predicted net load are obtained by the adaptive kernel density estimation under RES intermittency. Moreover, ECDM is extented to multi-energy forecast framework, attempting to increase interpretability of the net load predictions. Numerical experiments on a publicly available dataset demonstrate the superior forecasting performance of the proposed method compared to existing state-of-the-art approaches.
\end{abstract}



\begin{highlights}
\pagestyle{fancy}
\item An imputation-based conditional diffusion model is established to generate day-ahead net load scenarios guided by historical net load as well as weather and calendar information.
\item A novel feature engineering approach by weekly arrangement is introduced to improve forecast accuracy and an unconditional model is integrated into the proposed framework to ensure diversity in the generated scenarios.
\item The proposed method is extented to multi-energy forecasting, covering net load and its components, thereby offering comprehensive interpretability of the net load predictions. 
\end{highlights}

%% Keywords
\begin{keyword}
Renewable energy sources \sep probabilistic net load forecasting \sep conditional diffusion model\sep imputation mechanism\sep weekly arrangement \sep multi-energy forecast


\end{keyword}

\end{frontmatter}



\section{Introduction}
\label{sec:introduction}

As a consequence of the global transition to decarbonisation, the penetration ratio of distributed renewable energy sources (RES), in particular solar photovoltaic (PV) and wind power, is significantly increasing in modern power systems \cite{ref:renewable}. These innovative energy forms typically function as distributed sources within the energy market, contributing to reduced power outages and diminished network losses. Moreover, power grid users are transitioning from being mere consumers to 'prosumers'—entities that both consume and produce electricity, thus exhibiting dual characteristics of load and new energy externally.  In this novel paradigm, the conventional load forecasting challenge evolves into net load forecasting, wherein the forecast considers the actual load minus the output from RES \cite{ref:aggregated}. Accurate short-term net load forecasting is crucial for maintaining the balance between supply and demand, ensuring the future economic operation of power systems, and enhancing local consumption of renewable energy \cite{ref:optimal, ref:prosumers}.

The diversity and electrification of multiple types of load \cite{ref:efficient}, along with the intermittent and volatile nature of RES due to climate variability and natural conditions \cite{ref:battery, ref:management}, introduce uncertainties in net load and lead to a reduction in the precision of conventional deterministic net load forecasting \cite{ref:from}. The limitations of deterministic forecasting are well documented in contexts where increased uncertainty affects grid decision-making \cite{ref:data-driven, ref:probabilistic}. In contrast, probabilistic forecasting provides dispatchers with a more nuanced understanding of the inherent uncertainties, offering valuable insights into quantiles, intervals, and probability density functions \cite{ref:extreme}. 

Traditional probabilistic forecasting methods \cite{ref:probabilistic} can be broadly categorized into two groups: classical statistical techniques and machine learning approaches. Recently, researchers have increasingly focused on deep learning (DL), which boasts superior nonlinear fitting capabilities compared to traditional methods \cite{ref:ensemble}. DL Models such as long short-term memory (LSTM) network \cite{ref:lstm} and attention-based neural network \cite{ref:transformer} have demonstrated remarkable performance. In \cite{ref:density}, wind power is assumed to follow a beta mixture distribution, and an improved deep mixture density network is employed to predict the parameters of this distribution. However, assumptions regarding the predictive distribution may limit the scope of uncertainty quantification. In \cite{ref:enhancing}, an LSTM is trained using the pinball loss function to derive the quantiles of the predicted load. A prediction interval is constructed by residual bootstrapping in \cite{ref:explainable} using deeply-fused nets which is constructed from BiLSTM, CNN and other base networks. Additionally, the authors of \cite{ref:FPSeq2Q} propose the FPSeq2Q framework, which combines the attention mechanism with a deep neural network to achieve non-parametric conditional quantile prediction of net load. 

Methodologically, probabilistic forecasting aimed at  quantifying epistemic and aleatory uncertainties \cite{ref:uncertainty} through probability distribution can be categorized as follows \cite{ref:using}: a) generating multiple scenarios for deterministic models; b) developing probabilistic forecasting models; c) post-processing deterministic forecasts, and d) their combination. Most existing research has successfully demonstrated the efficacy of DL in probabilistic forecasting tasks along these lines. However, challenges in capturing correlations between uncertain variables and a lack of diversity in predictions can lead to model bias \cite{ref:review_scenario}. Furthermore, it is often more practical to make optimization decisions on a discrete scenario set rather than a probability distribution. Scenario forecasting, as a type of probabilistic forecasting, provides utility not only by informing operators about future uncertainties in the form of prediction intervals or quantile forecasts but also by generating plausible time series for early planning \cite{ref:short-term} to assess the impact of uncertainties. Various generative models have been proposed to produce diverse scenarios that encapsulate the correlative features of uncertain variables, including extreme events. Scenarios are sampled from the latent space of a prior probability distribution and then mapped to the actual probability distribution by generative models \cite{ref:generative}, effectively addressing epistemic uncertainty with aleatory uncertainty. \cite{ref:generative} reviews and contrasts several generative techniques, elucidating the underlying principles and interrelationships of each. Noting that conditions such as weather forecasts and deterministic predictions contribute to more accurate, controllable generation, \cite{ref:GAN} proposes a method for short-term load probability prediction using a conditional generative adversarial network (CGAN) for curve generation. \cite{ref:VAE} introduces a conditional variational autoencoder (CVAE) for short-term electrical load forecasting for residential and commercial accounts, trained on real-world consumption data. Nonetheless, the adversarial training process of GANs can introduce instability, while the performance of VAEs is restricted by the expressiveness of their variational posterior distribution \cite{ref:short-term}. A flexible approach utilizing Bernstein polynomial normalizing flows (NF) for conditional density forecasting of short-term load is proposed in \cite{ref:nf}, demonstrating advantages over other probabilistic techniques; however, this approach restricts the range of network structures and dimensions \cite{ref:nf-bad}.

As a generative model that has gained prominence in recent years, the denoising diffusion probabilistic model (DDPM) offers higher accuracy and stability than traditional generative models by utilizing a diffusion and denoising process anchored in Bayesian learning, thereby enhancing prediction uncertainty modeling \cite{ref:denoising}. Initially emerging in the field of image generation \cite{ref:diffusion}, diffusion models have gradually been extended to load forecasting \cite{ref:diffusion-based}, wind power forecasting \cite{ref:short-term, ref:diffusion-energy}, and other areas \cite{ref:latentdiff,ref:imputation}. Their exceptional ability to generate probabilistic distributions is poised to infuse new vigor into the research of probabilistic net load forecasting. In \cite{ref:short-term}, although a two-stage model is constructed, scenario generation still relies on deterministic forecasts, which increases computational complexity and does not fully exploit the probabilistic predictive advantages of diffusion models. Additionally, the model in \cite{ref:diffusion-energy} employs an overly simplistic condition embedding network, limiting its forecasting performance. Derivatives of the conditional diffusion model, such as the classifier-free guidance diffusion model \cite{ref:classifier-free} and the latent diffusion model \cite{ref:latent}, are capable of integrating data-related conditions using an attention-based neural network to generate content with more targeted intent. Studies such as \cite{ref:impainted} and \cite{ref:csdi} demonstrate the recovery of inpainted images using a diffusion-based imputation component with known accurate measurements, while \cite{ref:imputation} extends this approach to reconstruct power system measurements. Similarly, \cite{ref:diffusion-based} used a diffusion-based inpainting forecasting method to generate conditional deterministic electricity prediction. However, it did not investigate how DDPM performs in probabilistic forecasting. 

In light of conditional diffusion model with imputation, this paper proposes a novel enhanced conditional diffusion model (ECDM) for day-ahead probabilistic net load forecasting. This framework is designed to predict net load profiles using a probabilistic diffusion model while fully leveraging conditional guidance to generate realistic scenarios. The contributions of this study are summarized as follows:

\begin{enumerate}
\item An imputation-based conditional diffusion model is established to generate day-ahead net load scenarios. In this model, historical net load is treated as the known measurements to guide the generative process, while weather and calendar information are incorporated via the cross-attention mechanism.  
\item A novel feature engineering approach is introduced to improve forecast accuracy by leveraging the periodic characteristics of net load. Additionally, an unconditional model is integrated into the proposed framework to ensure diversity in the generated scenarios.
\item To enhance forecasting accuracy, an adaptive kernel density estimation method is employed. Moreover, the application of ECDM is extented to multi-energy forecasting, covering net load and its components, thereby offering comprehensive interpretability of the net load predictions. 
\end{enumerate}
Experimental results demonstrate that the proposed model outperforms several state-of-the-art methods and the effectiveness of the framework is validated through ablation experiments. 

The rest of this paper is organized as follows. Section \ref{sec:score} provides a brief description of the principle of the score-based diffusion model and how to incorporate conditional guidance. Section \ref{sec:probabi} elaborates the proposed probabilistic day-ahead net load forecasting framework and its components. Details about the experimental settings, results and discussion are presented in Section \ref{sec:case}. Finally, the paper is concluded in Section \ref{sec:Conclusion}.

% section introduction (end)

\section{Score-based Diffusion Model} \label{sec:score}
\subsection{Score-Based Generative Models}
Given a dataset $\mathbf{x} \in \mathbb{R}^n$ drawn from an unknown distribution $p(\bm{\textrm{x}})$, a stochastic differential equation (SDE) \cite{ref:solving} perturbs the data points with a stochastic process over a time horizon [0, 1] as
\begin{equation}
\label{eq_sde1}
\mathrm{d}\mathbf{x}_t = f(\mathbf{x}, t)\mathbf{x}_t \mathrm{d}t + g(t)\mathrm{d}\mathbf{w}_t, \quad t \in [0, 1],
\end{equation}
where $\mathbf{x}_t \in \mathbb{R}^n$ denotes a standard Wiener process (a.k.a., Brownian motion), $\mathbf{x}_t \in \mathbb{R}^n$ symbolizes the trajectory of random variables in the stochastic process, $t$ represents a timestep, $f(\mathbf{x}, t)$ and $g(t)$ are the drift and diffusion coefficients, respectively. 

Reversing the perturbation process described in Eq. \eqref{eq_sde1}, it is possible to start with a noise sample $\mathbf{x}_1 \sim p_1(\mathbf{x})$ and gradually remove the noise to obtain a data sample $\mathbf{x}_0 \sim p_0(\mathbf{x}) \equiv p(\mathbf{x})$. The reverse process is formulated in
\begin{equation}
  \label{eq_sde2}
  \mathrm{d}\mathbf{x}_t = [f(\mathbf{x}, t) - g^2(t)\nabla_{\mathbf{x}_t} \log p_t(\mathbf{x}_t)]\mathrm{d}t + g(t)\mathrm{d}\overline{\mathbf{w}}_t,
\end{equation}
where $\overline{\mathbf{w}}_t$ and ${\textrm{d}}t$ denote a standard Wiener process and an infinitesimal negative time step in the reverse-time direction. The quantity $\nabla_{\mathbf{x}_t} \log p_t(\mathbf{x}_t)$ is defined as the score function of $p_t(\mathbf{x}_t)$. Since the actual score fuction $\nabla_{\mathbf{x}_t} \log p_t(\mathbf{x}_t)$ is unknown, it's necessary to train a neural network $\boldsymbol{s_{\theta}}(\mathbf{x}, t)$ to approximate the actual one.

Given a dataset ${\mathbf{x}^{(1)}, \mathbf{x}^{(2)}, \dots , \mathbf{x}^{(N)}} \sim p(\mathbf{x})$, $\boldsymbol{s_{\theta}}(\mathbf{x}, t)$ can be trained 
with denoising score matching by solving the following objective function
\begin{equation}
  \label{eq_sde3}
  \begin{array}{c}
  \boldsymbol{\theta}^{*}=\mathop{\arg\min}\limits_{\boldsymbol{\theta}}\frac{1}{N}\sum_{i=1}^{N}
  \mathbb{E}_{t \sim \mathcal U [0, 1]}\mathbb{E}_{\mathbf{x}_t^{(i)} \sim p_{0t}(\mathbf{x}_t^{(i)}|\mathbf{x}^{(i)})}
  \\
  \left[\left \| \boldsymbol{s_{\theta}}(\mathbf{x}_t^{(i)}, t)-\nabla_{\mathbf{x}_t^{(i)}} \log p_{0t}(\mathbf{x}_t^{(i)}|\mathbf{x}^{(i)}) \right \|_2^2 \right],
  \end{array}
\end{equation}
where $\mathcal U [0, 1]$ denotes a uniform distribution over [0, 1]. After solving Eq. \eqref{eq_sde3}, an approximate score function $\boldsymbol{s_{\theta^{*}}}$ is obtained such that $\boldsymbol{s_{\theta^{*}}}(\mathbf{x},t) \approx \nabla_{\mathbf{x}_t^{(i)}} \log p_{0t}(\mathbf{x}_t^{(i)}|\mathbf{x}^{(i)})$. Then Eq. \eqref{eq_sde2} can be transformed into 
\begin{equation}
  \label{eq_sde4}
  % \begin{array}{c}
  \mathrm{d}\mathbf{x}_t = [f(\mathbf{x}, t) - g^2(t)\boldsymbol{s_{\theta^{*}}}(\mathbf{x}_t,t)]\mathrm{d}t + g(t)\mathrm{d}\overline{\mathbf{w}}_t.
  % \end{array}
\end{equation}

\subsection{Denoising Diffusion Probabilistic Model}
\label{sub:ddpm}

DDPM can be regarded as a discrete form of the score-based generative model. It generates new data samples by simulating a random diffusion process that gradually transforms random noise into a target data distribution. In particular, the diffusion model comprises two principal processes: a forward noise adding process and a reverse denoising process \cite{ref:denoising}.

The forward process, commonly known as the diffusion process, involves adding Gaussian noises to the original time series data $\boldsymbol{x}_{0} \sim p_0(\boldsymbol{x})$ over $T$ steps, until the data transforms into a standard Gaussian distribution $\boldsymbol{x}_{T} \sim \mathcal{N}(\boldsymbol{0},\boldsymbol{I})$. At each discrete timestep $t$, noise is introduced to the data $\boldsymbol{x}_t$ based on a predetermined intensity level as 
\begin{equation}
\label{ddpm1}
\begin{array}{c}
  \boldsymbol{x}_{t}=\sqrt{1-{\beta}_{t}} \boldsymbol{x}_{t-1}+\sqrt{\beta_{t}} \epsilon_{t-1},~~~\epsilon_{t-1} \sim \mathcal N(\boldsymbol{0}, \boldsymbol{I}),
\end{array}
\end{equation}
where ${\beta_{t} \in (0,1)}_{t=1}^{T}$ represents the variance of the noise schedule at each step and follows the condition $\beta_1 < \beta_2 < \dots < \beta_T$. The transformed probability is expressed as
\begin{equation}
\begin{aligned}
\label{ddpm2}
    q(\boldsymbol{x}_{t}|\boldsymbol{x}_{t-1})& =\mathcal{N}(\boldsymbol{x}_t;\sqrt{1-{\beta}_{t}} \boldsymbol{x}_{t-1},\beta_{t}\boldsymbol{I}).
\end{aligned}
\end{equation}
Since the whole forward process is a Markov chain, $\boldsymbol{x}_t$ in each step $t$ can be computed from $\boldsymbol{x}_0$ as $\boldsymbol{x}_{t}=\sqrt{\bar{\alpha}_{t}} \boldsymbol{x}_{0}+\sqrt{1-\bar{\alpha}_{t}} \epsilon$, where $ \bar{\alpha}_{t}=\prod_{k=1}^{t} \alpha_{k}$, ${\alpha}_{t}=1-\beta_t$, and $\epsilon$ is the equivalent Gaussian noise in the whole noise adding process.

The reverse denoising process initiates with an initial distribution of random Gaussian noise, $p(\boldsymbol{x}_{T})$, and progressively denoises over $T$ steps until the data follow the distribution of $\boldsymbol{x}_{0}$. The reverse conditional probability is expressed as
\begin{equation}
  \label{ddpm7}
  \scalebox{0.95}{$
  q(\boldsymbol{x}_{t-1}|\boldsymbol{x}_{t}, \boldsymbol{x}_0)=\mathcal{N}\left(\boldsymbol{x}_{t-1};\dfrac{1}{\sqrt{{\alpha}_{t}}} 
  \left(\boldsymbol{x}_{t}-\dfrac{\beta_t}{\sqrt{1-\bar{\alpha}_t}}\epsilon \right),\sigma_{t-1}^{2}\boldsymbol{I}\right),
  $}
\end{equation}
where $\sigma_{t-1}^{2}=\beta_t(1-\bar{\alpha}_{t-1})/(1-\bar{\alpha}_{t})$. The term $\epsilon$ represents the noise added during the diffusion process, which cannot be directly derived from $\boldsymbol{x}_t$. To estimate this quantity, a predictor $\epsilon_{\theta}(\boldsymbol{x}_{t},t)$ is employed. Consequently, The estimates of $\boldsymbol{x}_{t-1}$ can be obtained from $\boldsymbol{x}_t$ by
\begin{equation}
  \label{ddpm9}
    \boldsymbol{x}_{t-1}=\dfrac{1}{\sqrt{\alpha_{t}}}(\boldsymbol{x}_{t}-\dfrac{1-\alpha_{t}}{\sqrt{1-\bar{\alpha}_{t}}}\epsilon_{\theta}(\boldsymbol{x}_{t},t))
    + \sqrt{\dfrac{1-\bar{\alpha}_{t-1}}{1-\bar{\alpha}_{t}}\beta_t}z. 
\end{equation}
The diffusion models utilize a DL network to predict the noise $\epsilon_{\theta}(\boldsymbol{x}_{t},t)$ at each step $t$, thereby determining the predicted mean $\mu_{t-1}$ of $\boldsymbol{x}_{t-1}$. At this juncture, the score function is defined as $\boldsymbol{s_{\theta^{*}}}(\boldsymbol{x}_t,t)=-\frac{1}{\sqrt{1-\bar{\alpha}_t}}\epsilon_{\theta}(\boldsymbol{x}_{t},t)$, with its training objective given by
\begin{equation}
  \label{ddpm8}
  \begin{array}{c}
\boldsymbol{\theta}^{*}=\mathop{\arg\min}\limits_{\boldsymbol{\theta}}
  \mathbb{E}_{x_0\sim q_0(x),\epsilon \sim \mathcal{N}(0,I)}\left\| \epsilon-\epsilon_{\theta}(x_{t},t) \right\|_2^2.
\end{array}
\end{equation}
Typically, the UNet architecture based on convolutional operations is usually selected for the noise prediction model \cite{ref:denoising}.

\subsection{Classifier and Classifier-Free Guidance}
\label{cf}

Classifier guidance \cite{ref:diffusion} is employed to direct the generative process to conform to the input condition $\boldsymbol{y}$. According to Bayes' theorem, the conditioned probability $p(\boldsymbol{x}_t|\boldsymbol{y})$ is expressed as
$p(\boldsymbol{x}_t|\boldsymbol{y})=p(\boldsymbol{x}_t)p(\boldsymbol{y}|\boldsymbol{x}_t)/p(\boldsymbol{y})$. Consequently, the logarithmic gradient of $p(\boldsymbol{x}_t|\boldsymbol{y})$ produces a score function:
\begin{equation}
\label{cg2}
 \scalebox{0.96}{$
  \nabla_{\boldsymbol{x}_{t}}\log p_{\theta }(\boldsymbol{x}_{t}|\boldsymbol{y})=
  \underbrace{\nabla_{\boldsymbol{x}_{t}}\log p_{\theta }(\boldsymbol{x}_{t})}_{\rm{unconditional~score}}+
  \underbrace{\nabla_{\boldsymbol{x}_{t}}\log p_{\theta }(\boldsymbol{y}|\boldsymbol{x}_{t})}_{\rm{adversarial~gradient}}, 
  $}
\end{equation}
where $p_{\theta}$ represents a neural network approximation of the true probability distribution. Consequently, the conditional diffusion model with classifier guidance utilizes the gradient $\nabla_{\boldsymbol{x}_{t}}\log p_{\theta }(\boldsymbol{y}|\boldsymbol{x}_{t})$ to steer the score function's prediction towards the target distribution $p(\boldsymbol{x}_0|y)$. A parameter $\omega$ is typically used to modulate the scale of the guidance by
\begin{equation}
\label{cg3}
\begin{array}{l}
  \nabla_{\boldsymbol{x}_{t}}\log p_{\theta }(\boldsymbol{x}_{t}|\boldsymbol{y})=
  \nabla_{\boldsymbol{x}_{t}}\log p_{\theta }(\boldsymbol{x}_{t})+
  \omega\nabla_{\boldsymbol{x}_{t}}\log p_{\theta }(\boldsymbol{y}|\boldsymbol{x}_{t}).
\end{array}
\end{equation}

However, classifier guidance requires the additional training of an explicit classifier $\log p_{\theta }(\boldsymbol{y}|\boldsymbol{x}_{t})$, which incurs extra training costs and makes the generative results highly dependent on the classifier's accuracy. To circumvent this issue, classifier-free guidance \cite{ref:classifier-free} is implemented as an alternative conditional guidance approach. By substituting the weighted adversarial gradient in Eq. \eqref{cg3} with Eq. \eqref{cg2}, the following is obtained:
\begin{equation}
\label{cg4}
 \scalebox{0.94}{$
\nabla_{\boldsymbol{x}_{t}}\log p_{\theta }(\boldsymbol{x}_{t}|\boldsymbol{y})=
  \underbrace{(1-\omega)\nabla_{\boldsymbol{x}_{t}}\log p_{\theta }(\boldsymbol{x}_{t})}_{\rm{unconditional~score}}
   + \underbrace{\omega\nabla_{\boldsymbol{x}_{t}}\log p_{\theta }(\boldsymbol{x}_{t}|\boldsymbol{y})}_{\rm{conditional~score}}. $}
\end{equation}
In Eq. \eqref{cg4}, the guidance scale $\omega$ regulates the balance between the conditional and unconditional scores. Higher values of $\omega$ indicate a stronger dependence of the generative process on the condition $\boldsymbol{y}$. Classifier-free guidance employs an implicit classifier to integrate the condition, allowing the DDPM to train solely a noise predictor conditioned on $\bm{y}$.

\section{ECDM-Enabled Probabilistic Net Load Forecasting} \label{sec:probabi}

\subsection{Probabilistic Net Load Forecasting Framework}
\label{sub:framework}
\begin{figure}[!h]
\vspace{-0.1cm}
\centerline{\includegraphics[width=0.6\textwidth]{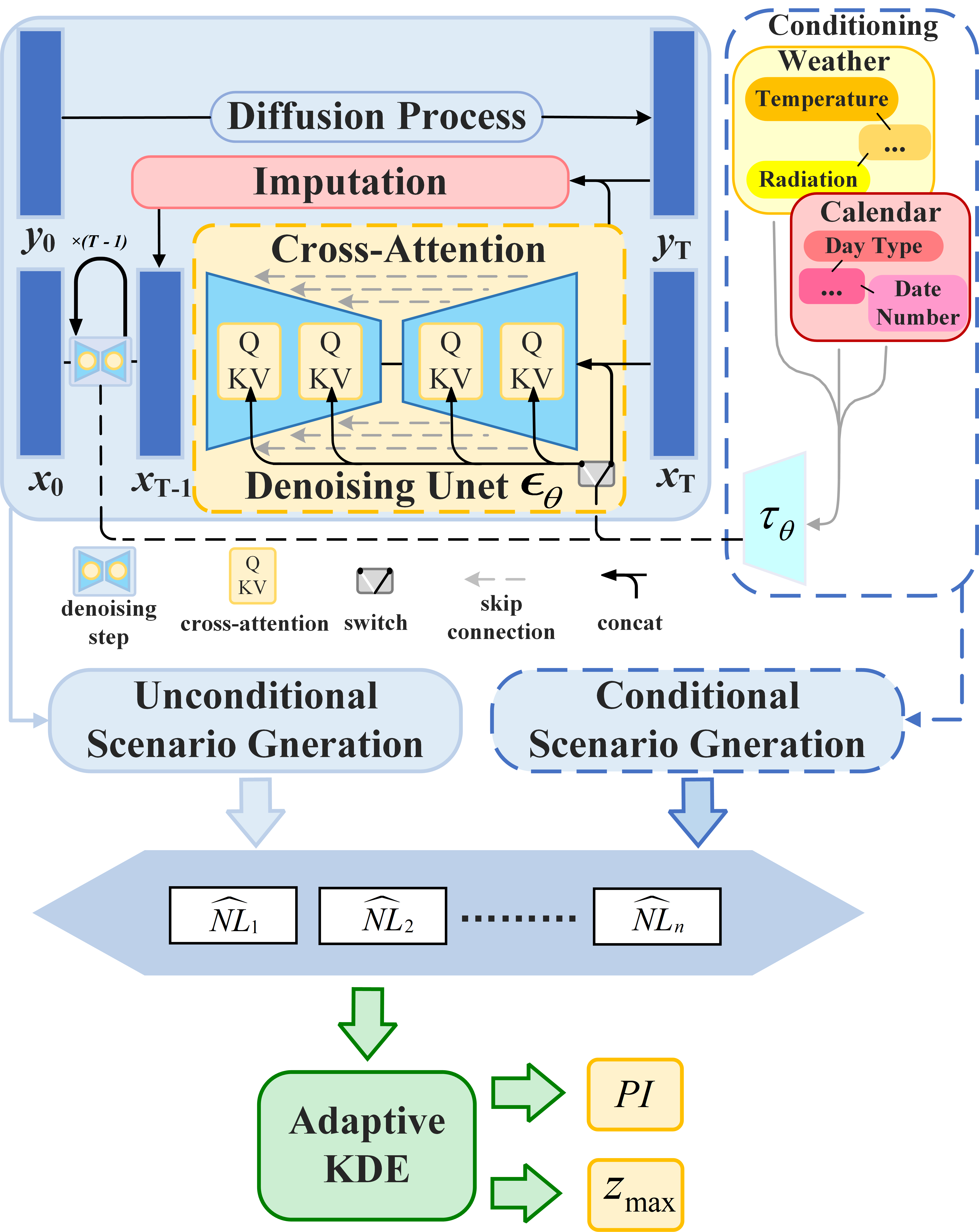}}
\caption{The overall structure of the proposed net load forecasting framework.}
\label{fig:framework}
\vspace{-0.2cm}
\end{figure}

The probabilistic net load forecasting framework, based on a cross-attention conditional diffusion model, is illustrated in Fig. \ref{fig:framework}. This framework comprises two main components: ECDM and adaptive kernel estimation. The ECDM is trained using net load data from both historical and forecast windows. During the training phase, the net load series is employed to train a noise predictor that leverages conditions such as weather and calendar information. This predictor is realized as a UNet, which integrates these conditions via cross-attention mechanisms. In the application phase, historical net load data is used in an imputation process for denoising, which facilitates the generation of multiple scenarios. The ECDM synthesizes all available conditions to steer the generation process but relies solely on diffusion models, excluding deterministic models. To enhance the diversity of scenarios, an unconditional model is also trained to produce partially unconditional samples. Following the generation of multiple scenarios, an adaptive kernel density estimation method is utilized to construct a more precise probabilistic interval. This interval is centered around the point of maximum probability.

\subsection{Feature Engineering by Weekly Arrangement}
\label{fe}
Traditional diffusion models in short-term energy forecasting typically generate scenario distributions $p_{\theta}(\boldsymbol{x}_0)$ based on deterministic results predicted by a neural network, where $\boldsymbol{x}_0$ represents the energy time series to be predicted. This paper proposes a method to forecast the net load using a single imputation stage. In short-term net load forecasting (SNLF), weather data $w$, such as numerical weather predictions (NWP), together with calendar information $c$ for the predicted day, are incorporated into the forecast framework, collectively expressed as $\boldsymbol{C}=[w_1, \cdots, w_m, c_1, \cdots, c_d]$. Factors such as temperature, solar radiation, and other NWP features, in addition to day type, weekday, and other calendar information, play an significant role in short-term scenario generation.

In addition to the mentioned features, the historical net load $\boldsymbol{y}_0=[NL_1, $ $NL_2, \cdots, NL_l]$ is considered, where $NL_k$ denotes the actual power at time $k$ in the historical window of length $l$. This historical net load is essential as it provides comprehensive time series data. The goal of SNLF is to predict the day-ahead net load profile $NL^f$, thus modifying the traditional forecasting distribution to $p_{\theta}(\boldsymbol{x}_0|\boldsymbol{y}_0, \boldsymbol{C})$. During the training stage, $\boldsymbol{x}_0=[NL_1^f, \cdots, NL_s^f]$ is provided, where $NL_k^f$ represents the actual net load at time $k$ within the forecast window, and $s$ is of the length window.

To construct time series that adhere closely to the same distribution for ECDM, the series $\boldsymbol{y}_0$ is rearranged based on weekly similarity. The previous six days are utilized as the historical window. As illustrated in Fig. \ref{fig:framework}, the historical window is arranged from Monday to Sunday, with the forecast window inserted among them: $\boldsymbol{y}_0=[NL_p, \cdots, NL_l, NL_1^f, \cdots, NL_s^f, NL_{1}, \cdots, NL_{p-1}]$, where $p$ indicates the first data point on Monday. Although this method changes the order of the historical window and introduces a discontinuity, it ensures that weekly data in net load profiles maintain a consistent distribution, since variations within a week are minor. This organizational method can be similarly adapted to monthly or annual characteristics.

\subsection{Cross-Attention Integrated UNet}
\label{ciu}
Classifier-free guidance is discussed in Subsection \ref{cf}. In this paper, the conditional model employs $\omega=1$ to fully utilize condition-based guidance for scenario generation. Weather and calendar data are provided as inputs to the neural network to predict the score function $\nabla_{\boldsymbol{x}_{t}}\log p_{\theta }(\boldsymbol{x}_{t}|\boldsymbol{y})$. The cross-attention mechanism, a widely adopted technique in DL, facilitates interaction between different inputs, enabling the model to effectively align contexts from various sources and better capture their correlations \cite{ref:latent}. This mechanism involves the computation of Query $Q$, Key $K$, and Value $V$ for two series $X_{1}$ and $X_{2}$ as follows:
\begin{equation}
\begin{array}{l}
    \text{CrossAttention}(X_{1},X_{2})=\text{Softmax}\left(\frac{QK^{T}}{\sqrt{d_{2}}}\right)V,
\end{array}
\end{equation}
where $Q = X_{1}W^{Q}$, $K = X_{2}W^{K}$, $V = X_{2}W^{V}$, and $W$s are parameter matrices; $d_2$ is a scaling factor.

The UNet architecture, originally developed for image segmentation tasks, can be employed to process time series data. It utilizes several convolutional neural networks (CNNs) for downsampling and upsampling. Cross-attention modules are integrated between the CNNs to incorporate additional conditions within the UNet. The UNet with cross-attention integration is depicted in Fig. \ref{fig:framework}, where it predicts noise as $\epsilon_{\theta}(\boldsymbol{x}_{t},t, \boldsymbol{C})$ based on the given conditions. Notably, several cross-attention modules are introduced in the intermediate layers of the UNet, which compute attention weights for features. This allows the model to focus on key features, thereby enhancing the UNet's adaptability to varying conditions.

\subsection{Imputation-Based Sampling Utilizing DDPM}
The forecasted net load, organized as described in Subsection \ref{fe}, can be viewed as containing missing entries, while the historical data can be considered an incomplete observation. This type of time series shares similarities with image inpainting, where the incomplete $\bm{y}_0$ guides the generation process by utilizing temporal correlations to complement the forecast positions $\bm{x}_0$. Let the index set of the historical net load be denoted as $\Omega$, and the index set for the forecast net load as $1-\Omega$. Thus, the historical data is represented as $\Omega \odot \boldsymbol{x}$, and the forecast data as $(1-\Omega) \odot \boldsymbol{x}$. The reconstruction process is outlined in Algorithm \ref{alg:alg1}.

\begin{figure}[!h]
\vspace{-0.3cm}
\centerline{\includegraphics[width=0.8\textwidth]{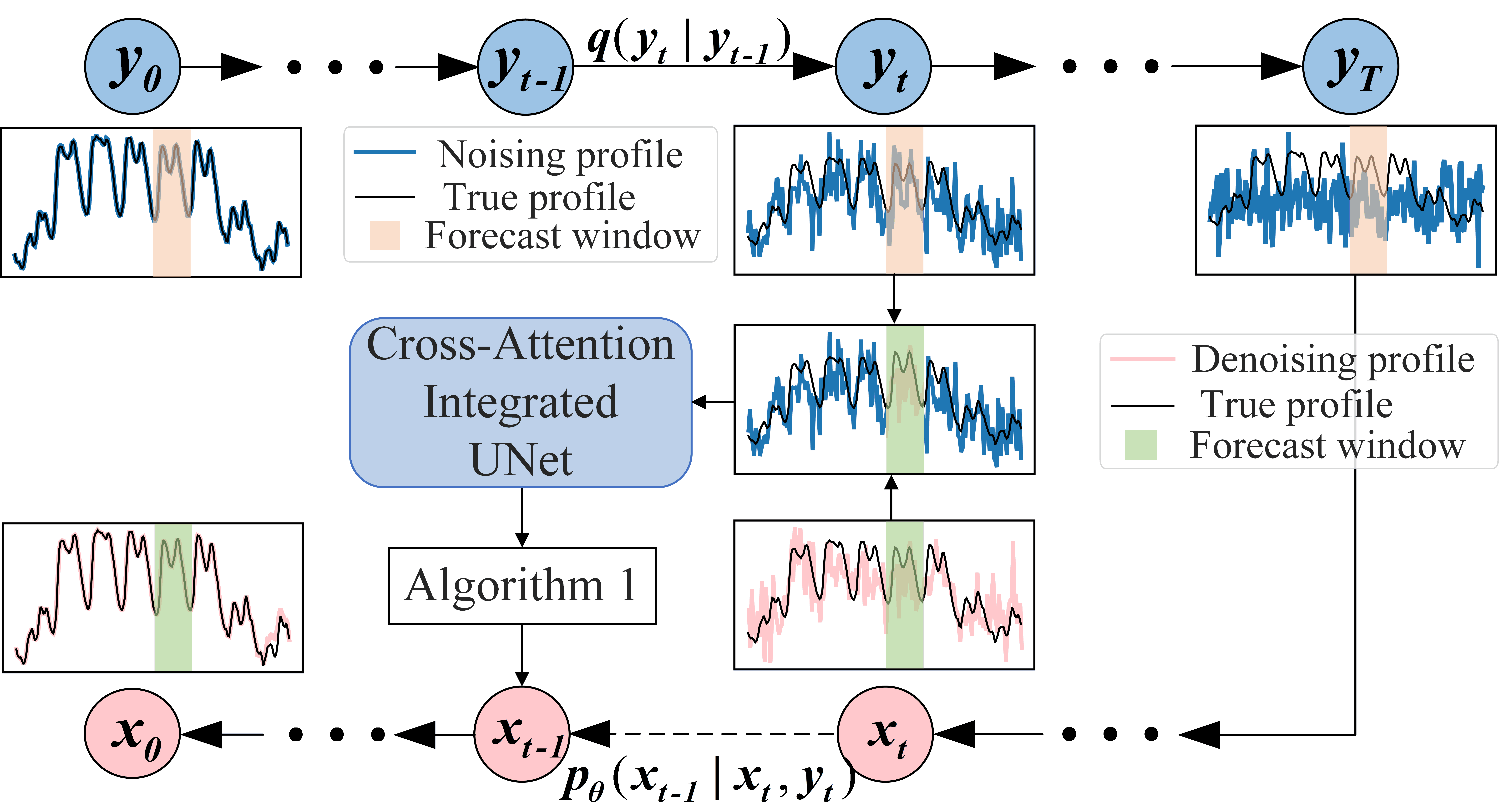}}
\caption{The scheme of imputation-based diffusion sampling process.}
\label{fig:imputation}
\vspace{-0.2cm}
\end{figure}

The training phase is discussed in Subsection \ref{sub:ddpm}, with slight modifications needed for the sampling process. At each step $t$, $\boldsymbol{x}_{t-1}$ is composed of noise-added $\boldsymbol{y}_0$ in the historical window and recovered data in the forecast window. Consequently, each iteration is guided by historical data, ultimately generating a forecasted net load that aligns with the historical window. The imputation scheme is illustrated in Fig. \ref{fig:imputation}, where the data array is described in Subsection \ref{fe}.

\begin{algorithm}
  \footnotesize
\caption{Imputation-based sampling for net load.}\label{alg:alg1}
\KwIn{The weather and calendar data $\boldsymbol{C}$, the number of scenarios $N$, the unconditional ratio $p_{uncond}$.}
\KwOut{Total scenario set $\boldsymbol{S}$}
Initialize $K \gets p_{uncond} \cdot N, M \gets N - K$\;
\For{$m \gets 1$ \textbf{to} $M$}{
    Initialize $\boldsymbol{x}_T \sim \mathcal{N}(0, \boldsymbol{I})$, $\alpha_t \gets 1 - \beta_t$, and $\bar{\alpha} \gets \prod_{t} \alpha_t$\;
    \For{$t \gets T$ \textbf{to} $1$}{
        Draw $\epsilon \sim \mathcal{N}(0, \boldsymbol{I})$ and $\boldsymbol{y}_{t-1} \gets \sqrt{\bar{\alpha}_t} \boldsymbol{y}_0 + \sqrt{1 - \bar{\alpha}_t} \epsilon$\;
        Draw $z \sim \mathcal{N}(0, \boldsymbol{I})$\;
        $\boldsymbol{x}_{t-1} \gets \dfrac{1}{\sqrt{\alpha_{t}}}(\boldsymbol{x}_{t}-\dfrac{1-\alpha_{t}}{\sqrt{1-\bar{\alpha}_{t}}}\epsilon_{\theta}(\boldsymbol{x}_{t},t,\bm{y}_0,\bm{C)})+\sqrt{\dfrac{1-\bar{\alpha}_{t-1}}{1-\bar{\alpha}_{t}}(1-\alpha_{t})}z$\;
        % $\boldsymbol{x}_{t-1} \gets \mu_{t-1} + \sigma_{t-1} z$\;
        $\boldsymbol{x}_{t-1} \gets \boldsymbol{x}_{t-1} \odot (1 - \Omega) + \boldsymbol{y}_{t-1} \odot \Omega$\;
    }
    Scenario generation: $\boldsymbol{s}^m \gets \boldsymbol{x}_0$\;}
    Conditional scenario set: $\boldsymbol{S}_{cond} = \left\{ \boldsymbol{s}^1, \boldsymbol{s}^2, \cdots, \boldsymbol{s}^M \right\}$\;
\For{$k \gets 1$ \textbf{to} $K$}{
    Initialize $\boldsymbol{x}_T \sim \mathcal{N}(0, \boldsymbol{I})$, $\alpha_t \gets 1 - \beta_t$, and $\bar{\alpha} \gets \prod_{t} \alpha_t$\;
    \For{$t \gets T$ \textbf{to} $1$}{
        Draw $\epsilon \sim \mathcal{N}(0, \boldsymbol{I})$ and $\boldsymbol{y}_{t-1} \gets \sqrt{\bar{\alpha}_t} \boldsymbol{y}_0 + \sqrt{1 - \bar{\alpha}_t} \epsilon$\;
        Draw $z \sim \mathcal{N}(0, \boldsymbol{I})$\;
        $\boldsymbol{x}_{t-1} \gets \dfrac{1}{\sqrt{\alpha_{t}}}(\boldsymbol{x}_{t}-\dfrac{1-\alpha_{t}}{\sqrt{1-\bar{\alpha}_{t}}}\epsilon_{\theta}(\boldsymbol{x}_{t},t,\bm{y}_0))+\sqrt{\dfrac{1-\bar{\alpha}_{t-1}}{1-\bar{\alpha}_{t}}(1-\alpha_{t})}z$\;
        % $\boldsymbol{x}_{t-1} \gets \mu_{t-1} + \sigma_{t-1} z$\;
        $\boldsymbol{x}_{t-1} \gets \boldsymbol{x}_{t-1} \odot (1 - \Omega) + \boldsymbol{y}_{t-1} \odot \Omega$\;
    }
    Scenario generation: $s^k \gets \boldsymbol{x}_0$\;
}
Unconditional scenario set: $\boldsymbol{S}_{uncond} = \left\{ \boldsymbol{s}^1, \boldsymbol{s}^2, \cdots, \boldsymbol{s}^K \right\}$\;
\textbf{Return:} Total scenario set: $\boldsymbol{S} = \left\{ \boldsymbol{S}_{cond}, \boldsymbol{S}_{uncond} \right\}$\;
\end{algorithm}

Weather information may be imprecise and epistemic uncertainty in models can introduce forecast errors. To address this, the condition-integrated diffusion model can enhance accuracy but may lack generative diversity. Therefore, this paper combines the conditional diffusion model with an unconditional one that retains the same structure as the former, but without cross-attention modules. This design enables the model to flexibly adjust the ratio of unconditional generation flexibly, thereby improving the accuracy and adaptability of the predictions. Referring to the sampling process outlined in Algorithm \ref{alg:alg1}, given the unconditional ratio $p_{uncond}$ and the total number of samples to be generated $n$, $p_{uncond} \times n$ unconditional samples are generated using noise-added data $\boldsymbol{x}_t$, while $n - p_{uncond} \times n$ conditional samples are generated with both $\boldsymbol{x}_t$ and condition $\boldsymbol{C}$. Ultimately, a set of $n$ samples is obtained by mixing these two types of data. The parameter $p_{uncond}$ serves as a hyperparameter to balance generative diversity and accuracy, thereby enhancing forecasting performance.

\subsection{Adaptive Kernel Density Estimation}
\label{akde}

ECDM is developed and effectively trained accordingly, leveraging historical net load data to inform an imputation approach. This method generates $n$ forecast net load profiles. Subsequently, the probabilistic forecast for the day-ahead net load is derived by using KDE. Within the methodology, $n$ sets of Gaussian noise are utilized to reconstruct $n$ sets of predicted profiles by applying reverse conditional and unconditional diffusion processes to the ECDM. For each time point $k \in \{1, \cdots, s\}$, a probability density function $f(z)$ is formed using the predicted values $\boldsymbol{Z}=\{z_{1},z_{2},\dots,z_{n}\}$estimated by the KDE, which can be written as
\begin{equation}
f(z)=\dfrac{1}{nh}\sum\limits_{i=1}^{n}K\left(\dfrac{z-z_{i}}{h}\right),
\end{equation}
where $K(\cdot)$ represents the kernel function, specifically the Gaussian kernel function in this application, and $h$ denotes the bandwidth estimated through the KDE method.

Subsequently, the cumulative distribution function $F(z)$ is computed to establish the lower and upper bounds, i.e., $q_{\underline{\alpha}}$ and $q_{\overline{\alpha}}$, of the confidence interval for a specified confidence level $\gamma$, defined as $q_{\alpha}=F(\alpha)^{-1}$ and $PI_\gamma=[q_{\overline{\alpha}},q_{\underline{\alpha}}]$, where $\gamma=\overline{\alpha}-\underline{\alpha}$ and $\overline{\alpha}$ and $\underline{\alpha}$ correspond to the upper and lower levels of the interval $PI_{\gamma}$.

Typically, a symmetrical interval is applied in KDE, fulfilling $\overline{\alpha}+\underline{\alpha}=1$. An adaptive interval construction methodology is introduced in this study. This approach calculates the maximum probability density point $z_{\text{max}}$ within $\boldsymbol{Z}$ based on $f(z)$, using it as a deterministic prediction measure and the center of the probabilistic interval. The interval boundaries are symmetrically aligned around this center. For simplicity, symmetrical data points are used to define the interval. Assuming $z_1 < z_2 < \dots < z_n$, the interval $PI_{\gamma}$ is determined by
\begin{equation}
c = \underset{i}{\arg\max} \, f(z_i),
\end{equation}
\begin{equation}
n_\gamma=\gamma \cdot n,
\end{equation}
\begin{equation}
PI_\gamma = 
\begin{cases}
[z_1, z_{1+n_\gamma}], & c - n_\gamma < 1, \\ 
[z_{n-n_\gamma}, z_{n}], & c + n_\gamma > n, \\
[z_{c-n_\gamma/2}, z_{c+n_\gamma/2}], & \text{otherwise.}
\end{cases}
\end{equation}
This adaptive KDE methodology constructs dynamic intervals centered on the point of maximum probability density, increasing the likelihood of encompassing the actual net load compared to traditional symmetrical intervals.

\subsection{Multi-Energy Forecasting Framework}
This subsection presents a framework designed to simultaneously predict load, RES, and net load. While numerous works forecast these components individually, relatively few address their simultaneous prediction. A diffusion model is developed by exploiting a series concatenated by load, RES, and net load as input. The UNet model, previously employed for 1-D series in subsection \ref{ciu}, is adapted here for 2-D series. 

The latter term in \eqref{cg3} poses a challenge for closed-form acquisition due to its dependence on time $t$ and the explicit relationship existing only between $\boldsymbol{y}$ and $\boldsymbol{x}_0$. The general form of the forward model is expressed as
\begin{equation}
\begin{array}{l}
    \boldsymbol{y} = \boldsymbol{\mathcal{A}}\left( \boldsymbol{x}_0 \right) + \boldsymbol{n}
    ,~ \boldsymbol{y}, \boldsymbol{n} \in \mathbb{R}^n, \boldsymbol{x}_0 \in \mathbb{R}^d,
  \end{array}
\end{equation}
where $\boldsymbol{\mathcal{A}}\left( \cdot \right): \mathbb{R}^d \mapsto \mathbb{R}^n$ is the forward measurement operator, and $\boldsymbol{n}$ represents measurement noise. The series fulfills a latent relationship, specifically, that the net load $\boldsymbol{NL}$ approximates the difference between load $\boldsymbol{L}$ and $\boldsymbol{RES}$. This relationship is reformulated as
\begin{equation}
  \label{F2}
\begin{array}{l}
  0 = \boldsymbol{\mathcal{A}}\left( [\boldsymbol{L}, \boldsymbol{RES}, \boldsymbol{NL}] \right) + \boldsymbol{n} = \boldsymbol{L} - \boldsymbol{RES} - \boldsymbol{NL} + \boldsymbol{n}.
  \end{array}
\end{equation}
For the given measurement model \eqref{F2} with $\boldsymbol{n} \sim \mathcal{N}(0, \sigma^2\boldsymbol{I})$, the approximation $p(\boldsymbol{y}|\boldsymbol{x}_{t}) \simeq p(\boldsymbol{y}|\hat{\boldsymbol{x}}_{0})$ holds true. By differentiating $p(\boldsymbol{y}|\boldsymbol{x}_{t})$ with respect to $\boldsymbol{x}_{t}$, the following expression is derived by
\begin{equation}
  \label{F3}
\begin{array}{l}
  \nabla_{\boldsymbol{x}_{t}}\log p(\boldsymbol{y}|\boldsymbol{x}_{t}) \simeq -\dfrac{1}{\sigma^2}\nabla_{\boldsymbol{x}_{t}} \left\| \boldsymbol{\mathcal{A}}\left( \hat{\boldsymbol{x}}_0(\boldsymbol{x}_{t}) \right) \right\|^2_2,
  \end{array}
\end{equation}
where $\hat{\boldsymbol{x}}_0:=\hat{\boldsymbol{x}}_0(\boldsymbol{x}_{t})$ indicates that the predicted $\boldsymbol{x}_0$ in DDPM Eq. \eqref{ddpm8} is a function of $\boldsymbol{x}_t$.

\begin{figure}[!h]
\vspace{-0.2cm}
\centerline{\includegraphics[width=0.7\textwidth]{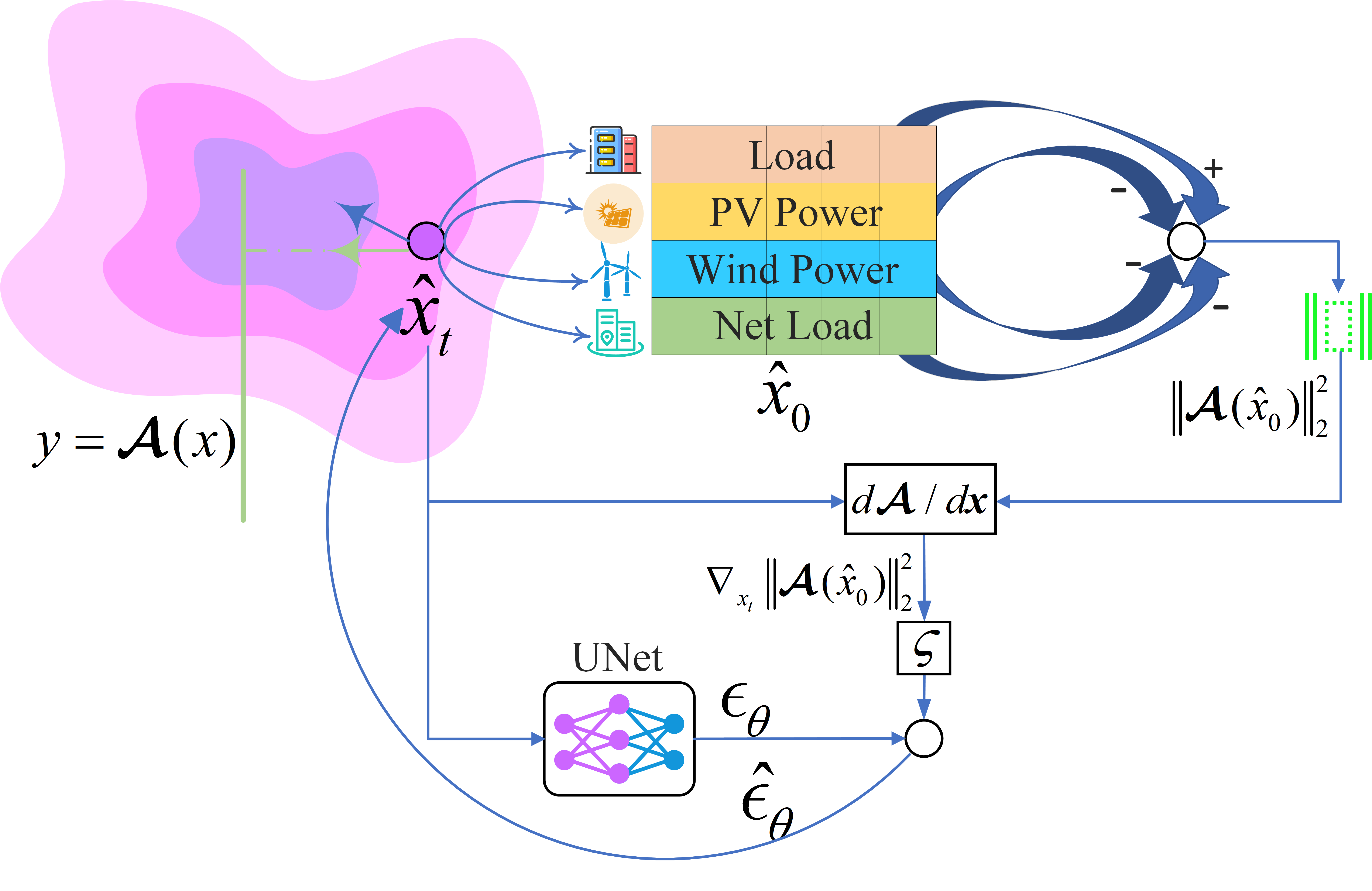}}
\caption{The schematic of multi-energy forecast.}
\label{fig:multi_scheme}
\vspace{-0.2cm}
\end{figure}

In conjunction with Eq. \eqref{cg3}, a simplified version of the equation is obtained to govern the generation process:
\begin{equation}
\label{F4}
\begin{array}{l}
  \nabla_{\boldsymbol{x}_{t}}\log p_{\theta }(\boldsymbol{x}_{t}|\boldsymbol{y})=
  \nabla_{\boldsymbol{x}_{t}}\log p_{\theta }(\boldsymbol{x}_{t})+
  \zeta\nabla_{\boldsymbol{x}_{t}} \left\| \boldsymbol{\mathcal{A}}\left( \hat{\boldsymbol{x}}_0(\boldsymbol{x}_{t}) \right) \right\|^2_2,
\end{array}
\end{equation}
where $\zeta$ serves as a scaling parameter to regulate the significance of the measurement guidance. The multi-energy forecast process, depicted in Fig. \ref{fig:multi_scheme}, is referred to as diffusion posterior sampling (DPS) \cite{ref:dps}. The score function grounded in DPS offers various directions for diffusion and steers the process closer to the path $y=\mathcal{A}(x)$.

\section{Case Study} \label{sec:case}
\subsection{Data Descriptions}
In this paper, experimental data comprising load, photovoltaic, and wind power measurements from the Baden-Württemberg region are utilized. These data are provided by TransnetBW, Germany \cite{ref:open}. The target synthetic net load is calculated as the difference between load and RES outputs. The TransnetBW datasets consist of separately reported load and generation measurements in 15-minute intervals, covering the period from 1st January 2015 to 31st December 2017. The training datasets include data from 2015 to 2016. For this study, 7 days from each season in 2017 are randomly chosen for validation, while the remaining days serve as test datasets. Weather and calendar features including wind speed, average temperature, time of day, and week type are sourced from a meteorological station in Baden-Württemberg and calendar records. Z-score normalization is applied to both net load power and meteorological data.

\subsection{Experimental Setup}

To illustrate the superior performance of the proposed model, a number of state-of-the-art probabilistic forecasting models, which have demonstrated reliable performance in the literature, are employed for comparison. These include generative models such as DDPM, GAN, VAE, and NF \cite{ref:diffusion-energy}. Additionally, other probabilistic forecasting models such as the improved quantile LSTM Network (IQLSTM) \cite{ref:iqlstm}, Bayes LSTM Network (BayesLSTM) \cite{ref:bayes}, and probabilistic model based on Dropout \cite{ref:dropout} are examined. The proposed method, ECDM, along with the comparative models, are trained using the same training set, with their hyperparameters determined through grid search over the validation datasets. All algorithms are executed using PyTorch on a server equipped with two Intel Xeon Gold 6248R 3.00 GHz CPUs, 128 GB RAM, and an NVIDIA RTX 3090 GPU. The hyperparameter settings employed in ECDM, as determined through validation, are detailed in Table \ref{tab_hyper}.

\begin{table}\footnotesize
\caption{Hyperparameter of the Proposed ECDM}
\label{tab_hyper}
\centering
\renewcommand{\arraystretch}{1.0}
\setlength\tabcolsep{1.1em}
% \resizebox{1.0\linewidth}{!}{
\begin{tabular}{ccccccccccc}
\hline
\textbf{Hyperparameter} & {\textbf{Definition}} & {\textbf{Value}}\\
\hline
$(\beta_0, \beta_T)$ & Variance of noise\% & (1e-4, 0.02)\\
$T$  & Diffusion step & 50\\
$l_r$ & Learning rate & 5e-4\\
$b$ & Batch size & 64 \\
$e$ & Epoch & 500 \\
$N$  & Number of scenarios & 100 \\
$p_{uncond}$ & Unconditional ratio & 0.28 \\
$d$ & Depth of UNet & 6 \\
$h$ & Attention head & 8 \\
\hline
\end{tabular}
% }
\end{table}

\subsection{Evaluation Metrics}
Standard evaluation metrics are used to assess the forecasting performance of the methods examined. These include mean absolute percentage error (MAPE) for deterministic forecasts and the average coverage error (ACE), the prediction interval average width (PIAW) alongside the Winkler score for probabilistic interval forecasts. MAPE quantifies the absolute percentage difference between the actual and predicted net loads. Given the actual net load $NL_{i}$ and the predicted net load $\hat{NL}_{i}$, MAPE is calculated as 
\begin{equation}
    MAPE = \frac{1}{m} \sum_{i=1}^{m} \frac{|NL_{i} - \hat{NL}_{i}|}{NL_{i}},
\end{equation}
where $\hat{NL}_{i}$ represents the deterministic net load prediction.

The average coverage error (ACE) and the prediction interval average width (PIAW) are employed to assess reliability and accuracy. ACE and PIAW which indicate the reliability and sharpness of prediction intervals, are defined by:
\begin{equation}
    c_i = 
    \begin{cases} 
    1, & NL_{i} \in [q_{\underline{\alpha}}^{i}, q_{\overline{\alpha}}^{i}],\\[3mm]
    0, & NL_{i} \notin [q_{\underline{\alpha}}^{i}, q_{\overline{\alpha}}^{i}],
    \end{cases}
\end{equation}
\begin{equation}
    ACE = \frac{1}{m} \sum_{i=1}^{m} c_i - \gamma,
\end{equation}
\begin{equation}
    PIAW = \frac{1}{m} \sum_{i=1}^{m} \left(q_{\overline{\alpha}}^{i} - q_{\underline{\alpha}}^{i}\right),
\end{equation}
where $\gamma$ is the nominal confidence level of the interval; $\underline{\alpha}$ and $\overline{\alpha}$ represent the quantile levels for the interval bounds, with $q_{\underline{\alpha}}^{i}$ and $q_{\overline{\alpha}}^{i}$ being the quantiles of the $i$th sample's bounds. Ideally, $ACE \approx 0$.

Additionally, the Winkler score serves as a comprehensive metric for probabilistic forecasting, assessing both accuracy and interval width by
\begin{equation}
    Winkler =
    \begin{cases} 
    q_{\overline{\alpha}}^{i} - q_{\underline{\alpha}}^{i} + \frac{2}{\alpha}(q_{\underline{\alpha}}^{i} - NL_{i}), & NL_{i} < q_{\underline{\alpha}}^{i},\\[3mm]
    q_{\overline{\alpha}}^{i} - q_{\underline{\alpha}}^{i} + \frac{2}{\alpha}(NL_{i} - q_{\overline{\alpha}}^{i}), & NL_{i} > q_{\overline{\alpha}}^{i},\\[3mm]
    q_{\overline{\alpha}}^{i} - q_{\underline{\alpha}}^{i}, & \text{otherwise.}
    \end{cases}
\end{equation}
The Winkler score is useful for quantile and interval forecasts as a comprehensive assessment of reliability and sharpness. A lower Winkler score indicates better probabilistic estimation.

\subsection{Deterministic and Probabilistic Forecasting Results}
The performance of the proposed ECDM is evaluated in both deterministic and probabilistic forecasting and compared with other existing probabilistic methods. The deterministic performance is assessed using the 50th percentile forecasts from each model, while the probabilistic capability is quantified through symmetric prediction intervals. The input features include historical net load, weather conditions, and calendar information.

Table \ref{tab_models} presents the forecasting results across four seasons, with the two best-performing models highlighted in bold for each metric. The proposed ECDM achieves the lowest MAPE of 7.19\%, along with the second lowest ACE and Winkler score of 0.80\% and 2598.67, respectively. These results indicate that ECDM produces accurate scenario forecasts with reliable confidence intervals. Additionally, its interval width ranks fourth at 1589.84, demonstrating good sharpness.

\begin{table}
\footnotesize
\caption{Forecast Performance Comparison Between Proposed Model and Contrast Models for Different Seasons}
\label{tab_models}
\centering
\renewcommand{\arraystretch}{1.0}
\setlength\tabcolsep{0.8em}

% 第一部分：左边的模型
\begin{tabular}{ccccccc}
\hline
\multicolumn{2}{c}{} & \textbf{DDPM} & \textbf{GAN} & \textbf{VAE} & \textbf{NF} \\
\hline
\multirow{4}{*}{Spring} & MAPE & 14.78\% & 8.60\% & 11.64\% & 9.33\% \\
                        & ACE  & 2.04\% & \textbf{-0.90\%} & -6.27\% & -2.93\% \\
                        & PIAW & 3082.36 & 2329.42 & 1863.56 & 2064.73 \\
                        & Winkler & 5166.09 & 4006.89 & 3659.32 & 3525.99 \\
\hline
\multirow{4}{*}{Summer} & MAPE & 21.58\% & 9.38\% & 17.93\% & 11.49\% \\
                        & ACE  & 1.71\% & \textbf{1.26\%} & -16.83\% & 4.78\% \\
                        & PIAW & 3304.62 & 2186.10 & 1776.77 & 2073.98 \\
                        & Winkler & 5762.30 & 3307.23 & 4464.41 & \textbf{2932.52} \\
\hline
\multirow{4}{*}{Autumn} & MAPE & 15.98\% & 8.46\% & 13.74\% & 10.59\% \\
                        & ACE  & 8.31\% & \textbf{-1.30\%} & -9.20\% & 6.69\% \\
                        & PIAW & 3332.91 & 2033.26 & 1721.68 & 2012.62 \\
                        & Winkler & 4820.55 & 3152.79 & 3826.76 & 2545.81 \\
\hline
\multirow{4}{*}{Winter} & MAPE & 15.49\% & 9.23\% & 13.21\% & 10.70\% \\
                        & ACE  & -1.46\% & \textbf{-3.56\%} & -17.17\% & -6.27\% \\
                        & PIAW & 3076.65 & 2133.84 & 1848.93 & 2121.58 \\
                        & Winkler & 5220.51 & 5111.14 & 4964.94 & 4370.42 \\
\hline
\multirow{4}{*}{Total}  & MAPE & 16.95\% & 8.92\% & 14.13\% & 10.53\% \\
                        & ACE  & 2.62\% & -1.78\% & -12.38\% & \textbf{0.57\%} \\
                        & PIAW & 3198.39 & 3898.14 & 1863.56 & 2064.73 \\
                        & Winkler & 5242.23 & 3611.24 & 4231.10 & 3346.81 \\
\hline
% \end{tabular}

% \vspace{0.6em}

% % 第二部分：右边的模型
% \begin{tabular}{ccccccc}
\hline
\multicolumn{2}{c}{} & \textbf{IQLSTM} & \textbf{BayesLSTM} & \textbf{Dropout} & \textbf{ECDM} \\
\hline
\multirow{4}{*}{Spring} & MAPE & 7.16\% & \textbf{5.70\%} & 8.79\% & \textbf{6.40\%} \\
                        & ACE  & -4.15\% & -7.13\% & \textbf{-1.10\%} & 2.60\% \\
                        & PIAW & \textbf{1290.97} & \textbf{904.25} & 1400.77 & 1579.84 \\
                        & Winkler & 2948.24 & \textbf{2505.95} & 3058.86 & \textbf{2437.21} \\
\hline
\multirow{4}{*}{Summer} & MAPE & 11.68\% & \textbf{7.16\%} & 12.74\% & \textbf{8.43\%} \\
                        & ACE  & -8.61\% & -7.81\% & -7.13\% & \textbf{-0.90\%} \\
                        & PIAW & 1364.60 & \textbf{878.48} & \textbf{1308.18} & 1718.80 \\
                        & Winkler & 3548.65 & 3654.73 & 3388.10 & \textbf{2717.57} \\
\hline
\multirow{4}{*}{Autumn} & MAPE & 6.89\% & \textbf{5.94\%} & 10.32\% & \textbf{5.50\%} \\
                        & ACE  & 5.85\% & \textbf{-3.76\%} & -7.39\% & 9.84\% \\
                        & PIAW & 1313.16 & \textbf{835.00} & \textbf{1146.93} & 1451.89 \\
                        & Winkler & \textbf{2085.38} & 2248.41 & 2951.20 & \textbf{1711.09} \\
\hline
\multirow{4}{*}{Winter} & MAPE & 8.42\% & \textbf{6.12\%} & 8.56\% & \textbf{8.34\%} \\
                        & ACE  & -9.72\% & -12.81\% & \textbf{-5.53\%} & -8.21\% \\
                        & PIAW & \textbf{1323.06} & \textbf{902.25} & 1462.78 & 1608.58 \\
                        & Winkler & 3661.62 & \textbf{2827.87} & \textbf{3003.89} & 3517.61 \\
\hline
\multirow{4}{*}{Total}  & MAPE & 8.54\% & \textbf{6.23\%} & 10.09\% & \textbf{7.19\%} \\
                        & ACE  & -4.19\% & -7.91\% & -7.16\% & \textbf{0.80\%} \\
                        & PIAW & 1322.95 & \textbf{880.13} & \textbf{1330.47} & 1589.84 \\
                        & Winkler & 3064.61 & \textbf{2511.95} & 3099.93 & \textbf{2598.67} \\
\hline
\end{tabular}
\end{table}

Compared to other generative models, including DDPM, GAN, VAE, and NF, ECDM exhibits superior overall performance by achieving a more precise deterministic forecast while maintaining a well-balanced prediction interval. Other generative approaches tend to generate wider intervals under the same confidence level, leading to higher Winkler scores. Among all models, BayesLSTM and ECDM consistently rank as the top performers across all four seasons, demonstrating stable forecasting capabilities.

For methods based on deterministic LSTM, including IQLSTM, BayesLSTM, and Dropout, lower MAPEs suggest strong deterministic forecasting abilities. In particular, BayesLSTM excels in both deterministic prediction and interval width. However, these models tend to produce excessively narrow confidence intervals, resulting in negative ACE values of -4.19\%, -7.91\%, and -7.16\%, respectively, which implies a lack of reliability in practical applications. These findings suggest that LSTM-based models may suffer from overfitting while failing to provide robust probabilistic forecasts. In contrast, ECDM effectively balances probabilistic generation capability, accuracy, and interval reliability.

\subsection{Analysis of Seasonal and Interval Forecasting Characteristics}
\begin{figure}[!h]
\vspace{-0.2cm}
\centerline{\includegraphics[width=0.7\textwidth]{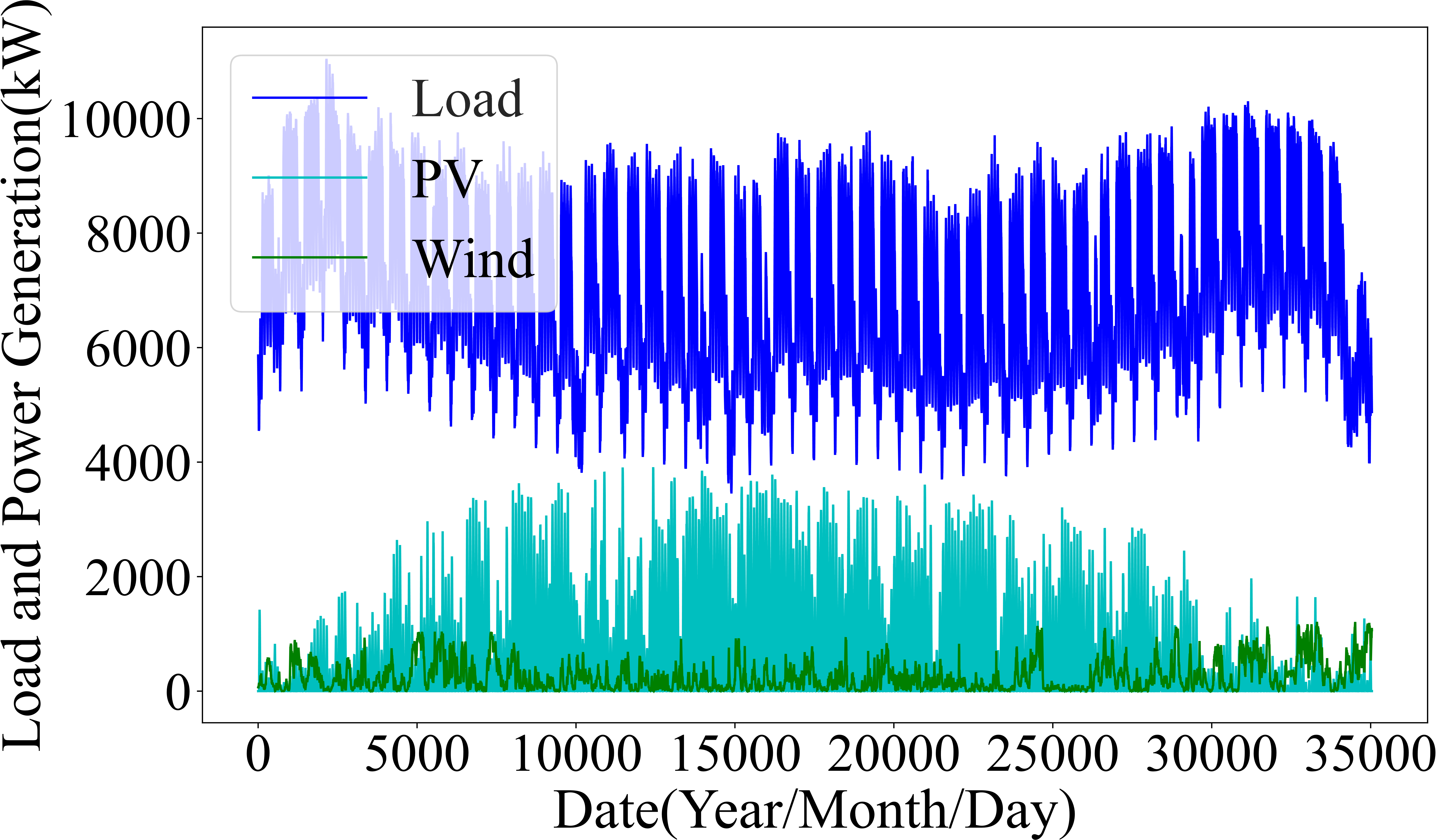}}
\caption{Actual load and RES power in 2017.}
\label{fig:season}
\vspace{-0.2cm}
\end{figure}

Forecasting performance varies across seasons due to fluctuations in RES. As depicted in Fig. \ref{fig:season}, RES exhibits higher volatility in summer and winter in 2017, whereas it remains more stable in spring and autumn. Consequently, probabilistic forecast accuracy tends to degrade in summer and winter across all models, including ECDM.

\begin{figure*}[!h]
\vspace{-0.2cm}
\centerline{\includegraphics[width=1.0\textwidth]{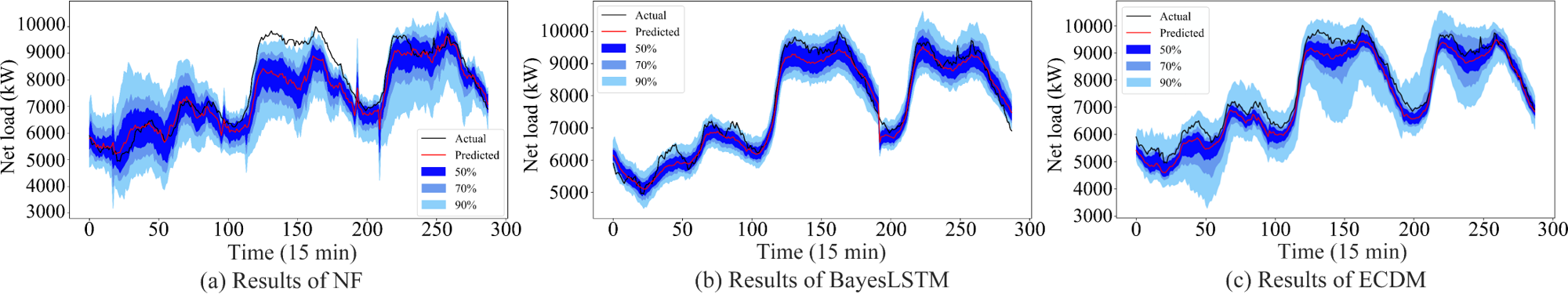}}
\caption{Probabilistic forecasting result of NF, BayesLSTM and ECDM under in three days.}
\label{fig:prediction}
\vspace{-0.2cm}
\end{figure*}

Fig. \ref{fig:prediction} showcases the probabilistic forecasts for net load over 96 points in a three-day period using NF, BayesLSTM, and ECDM. The actual net load is depicted by the black curve, while the red curve indicates the predicted net load, with blue-shaded regions representing confidence intervals at 50\%, 70\%, and 90\% levels. ECDM demonstrates a more compact interval compared to NF, providing improved coverage over BayesLSTM. Specifically, NF results in excessively wide intervals and less accurate forecasts, whereas BayesLSTM produces narrower intervals but fails to capture finer variations in the net load curve. ECDM strikes a favorable balance, providing higher reliability, which is crucial for power system operations, although there is still a need for further enhancements in interval sharpness. Forecast intervals also vary throughout the day. During nighttime, RES fluctuations are minimal, resulting in narrower intervals. In contrast, during the midday hours, RES generation increases, leading to wider intervals. Notably, the maximum probability density points closely match the actual net load values, and the forecast intervals successfully capture most variations while dynamically adjusting their widths. This capability enhances the elasticity of probabilistic forecasting, making the model more adaptable to net load fluctuations.

\begin{figure*}[!h]
 \vspace{-0.1cm}
\centerline{\includegraphics[width=1.0\textwidth]{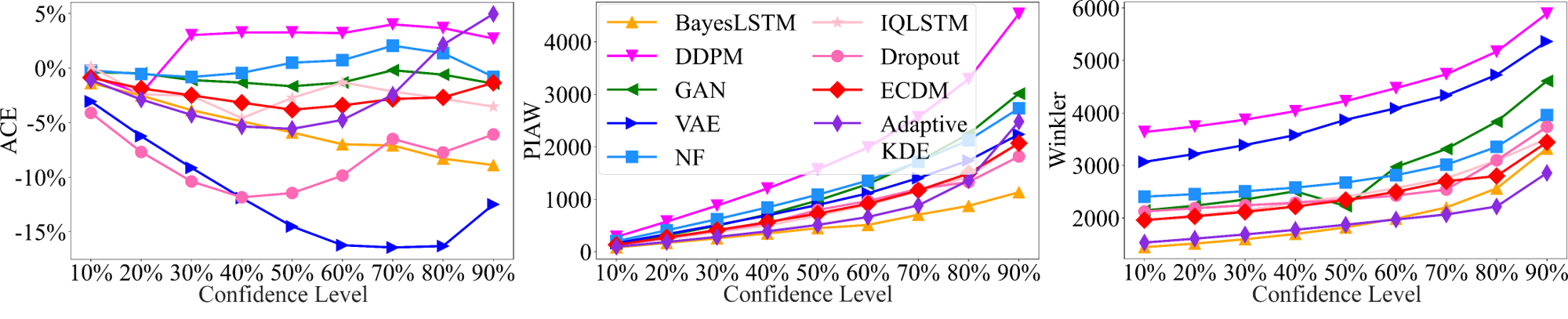}}
\caption{Probabilistic forecasting results of ECDM under various confidence levels, evaluated using ACE, PIAW, and Winkler score.}
\label{fig:confidence}
\vspace{-0.4cm}
\end{figure*}

Fig. \ref{fig:confidence} evaluates the reliability and sharpness performance of ECDM across various confidence levels $\alpha$. It is evident that ECDM achieves an ACE score close to zero, indicating higher reliability. Moreover, its ACE remains stable across different confidence levels, and the PIAW of ECDM increases with higher $\alpha$. In contrast, other methods exhibit lower reliability, with greater fluctuations in performance and a trade-off between interval width and quality as $\alpha$ changes. Notably, the ACE for BayesLSTM, VAE, and Dropout may drop below -5\%, demonstrating their lack of reliability. ECDM, however, effectively utilizes historical net load, weather, and calendar information, balancing reliability and accuracy in interval forecasting to achieve superior overall performance.

An adaptive KDE method is also incorporated to refine forecasting accuracy. As described in Section \ref{akde}, this approach constructs an asymmetric confidence interval centered on the most likely net load value. Fig. \ref{fig:confidence} (violet lines) illustrates the advantages of the proposed ECDM over standard KDE, demonstrating narrower confidence intervals and reduced Winkler scores while maintaining high coverage probability. The calculated MAPE values of 6.86\% further confirm that the use of the maximum probability data point improves forecast accuracy and increases the coverage of the actual values.

\subsection{Weekly Arrangement vs Daily Arrangement}
\label{result:data arrangement}
\begin{figure}[!h]
\vspace{-0.2cm}
\centerline{\includegraphics[width=0.7\textwidth]{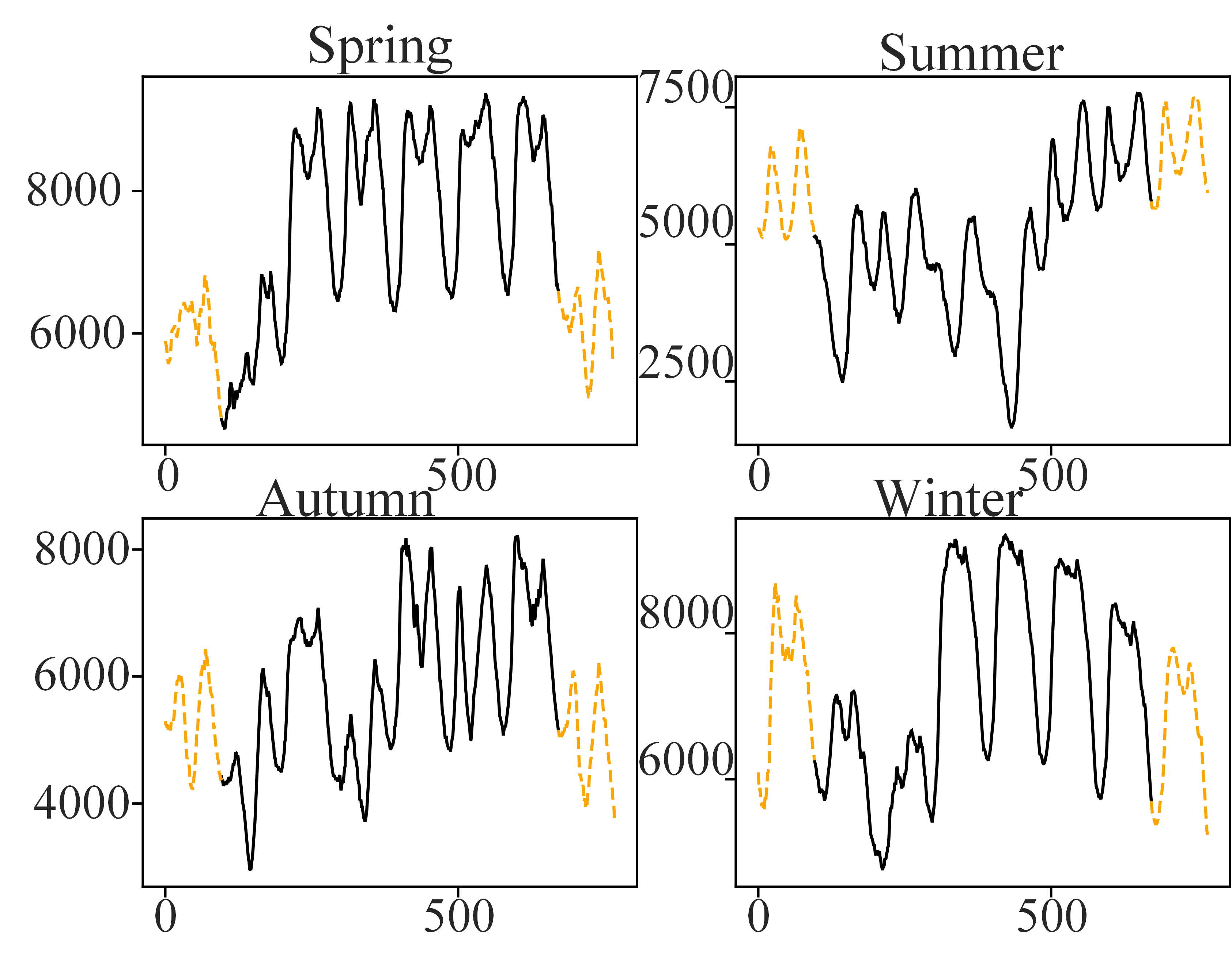}}
\caption{Typical net load data in 8 days (the orange dashed line represents the forecast day and the the same day last week, the black line represents data of the other days).}
\label{fig:week}
\vspace{-0.2cm}
\end{figure}

A novel data arrangement method is introduced in Subsection \ref{fe}, treating historical data as an image to leverage weekly patterns and enhance prediction accuracy, thereby moving beyond the conventional chronological order. By incorporating weekly similarities, the proposed method enables ECDM to generate more precise forecasts. Table \ref{tab_weekly} compares the deterministic and probabilistic results of the traditional chronological (daily) arrangement and the proposed weekly arrangement. The latter consistently outperforms the former throughout the year, achieving approximately 20\% relative improvement in PIAW and an ACE closer to zero, leading to superior overall performance. Fig. \ref{fig:week} presents representative 8-day data, demonstrating that in summer and autumn, daily load patterns closely resemble those of the same weekday in the previous week. This observation aligns with Fig. \ref{fig:season}, where seasonal load trends exhibit similar behavior. Consequently, the weekly arrangement effectively reduces prediction interval width and mitigates uncertainty in probabilistic forecasting. By fully exploiting weekly characteristics, the proposed method enhances scenario generation and improves forecasting accuracy.

\begin{table}[h]
\footnotesize
\caption{Deterministic and Probabilistic Forecast Results For Chronological Arrangement vs Weekly Arrangement}
\label{tab_weekly}
\centering
\renewcommand{\arraystretch}{1.0}
\setlength\tabcolsep{1.45em}
% \resizebox{1.0\linewidth}{!}{
\begin{tabular}{ccccccccccc}
\hline
\multicolumn{2}{c}{} & \textbf{Chronological} & {\textbf{Weekly}}\\
\hline
\multirowcell{4}{Spring} & MAPE & 5.69\% & 6.40\%\\
                          & ACE  & 5.12\% & 2.60\%\\
                          & PIAW & 1991.30 & 1579.84\\
                          & Winkler & 2654.00 & 2437.21 \\
\hline
\multirowcell{4}{Summer} & MAPE & 9.97\% & 8.43\% \\
                          & ACE  & -0.32\% & -0.90\% \\
                          & PIAW & 2199.10 & 1718.80 \\
                          & Winkler & 3148.09 & 2717.57 \\
\hline
\multirowcell{4}{Autumn} & MAPE & 6.45\% & 5.50\% \\
                          & ACE  & 10.79\% & 9.84\% \\
                          & PIAW & 2077.38 & 1451.89 \\
                          & Winkler & 2318.03 & 1711.09 \\
\hline
\multirowcell{4}{Winter} & MAPE & 6.50\% & 8.34\% \\
                          & ACE  & 1.54\% & -8.21\% \\
                          & PIAW & 2046.75 & 1608.58 \\
                          & Winkler & 3002.41 & 3517.61 \\
\hline
\multirowcell{4}{Total} & MAPE & 7.15\% & 7.19\% \\
                          & ACE  & 4.27\% & 0.80\% \\
                          & PIAW & 2078.54 & 1589.84 \\
                          & Winkler & 2781.31 & 2598.67 \\
\hline
\end{tabular}
% }
\end{table}

\subsection{Different Unconditional Ratio}
\begin{figure}[!h]
\vspace{-0.2cm}
\centerline{\includegraphics[width=0.7\textwidth]{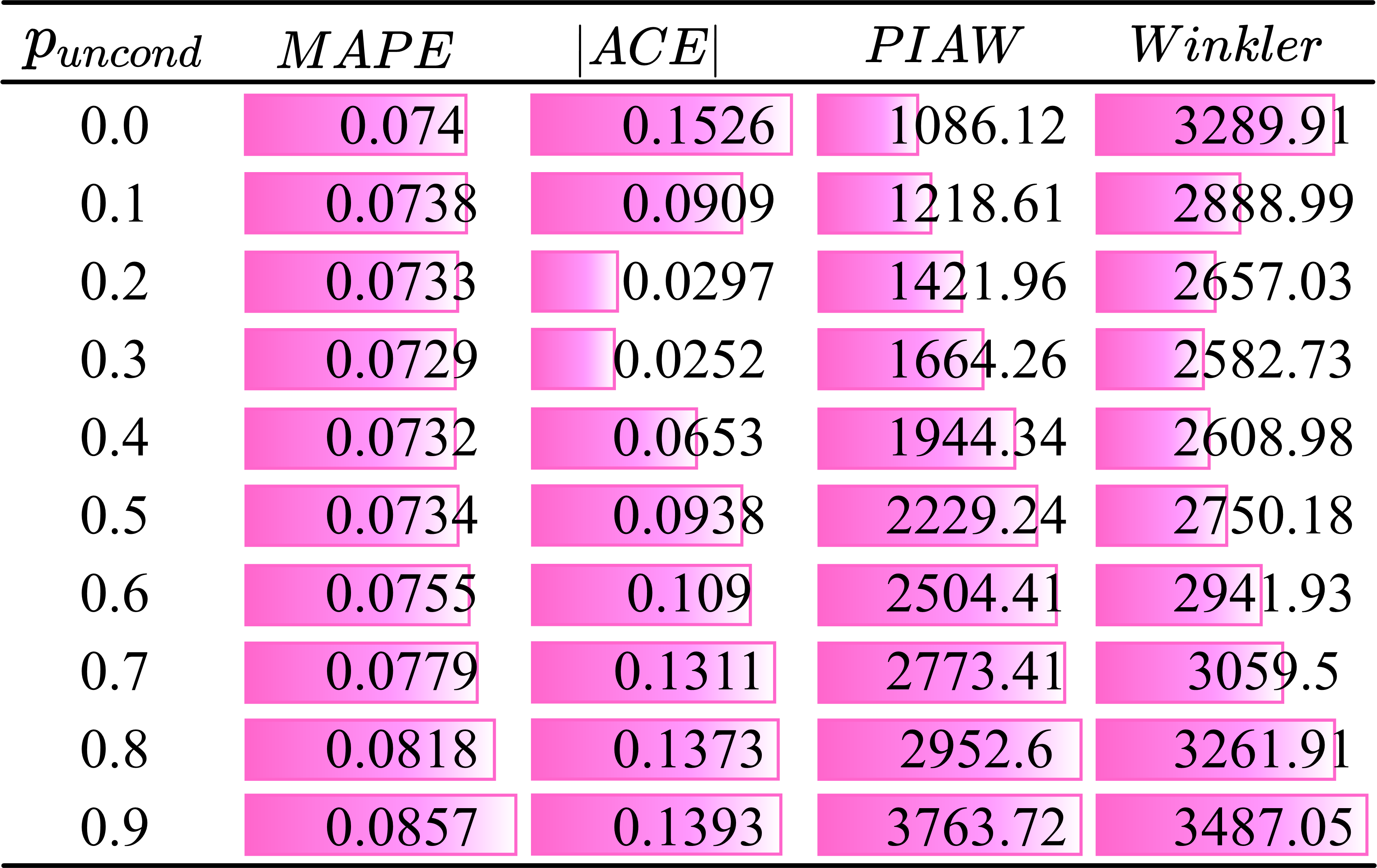}}
\caption{Net load forecasting performance across different $p_{uncond}$.}
\label{fig:p_uncond}
\vspace{-0.2cm}
\end{figure}

After showcasing the robust probabilistic forecasting capability of the ECDM, this subsection explores the impact of the unconditional ratio $p_{uncond}$ in Algorithm \ref{alg:alg1} on the accuracy of probabilistic scenario generation. An 80\% confidence level interval prediction is conducted using ECDM with $p_{uncond}$ set to $[0.0, 0.1, 0.2, \ldots, 0.9]$. Evaluation metrics for various $p_{uncond}$ values are depicted in Fig. \ref{fig:p_uncond}. The absolute value of ACE decreases starting from $p_{uncond}=0.0$, reaching a minimum at $p_{uncond}=0.3$, while PIAW consistently increases, underscoring the value of blending conditional and unconditional scenarios. Conditional generation, informed by weather and calendar data, yields more precise predictions but with less actual net load coverage. In contrast, unconditional generation enhances scenario diversity at the cost of interval quality and width. A balance between forecasting accuracy and diversity can be tailored to the requirements of the power system operator. At $p_{uncond}=0.3$, optimal performance with $|ACE|=0.0252$, $PIAW=1664.26$, and $Winkler=2582.73$ is achieved. Operators prioritizing reliability and wider intervals may opt for a larger $p_{uncond}$.

\begin{figure}[!h]
\vspace{-0.1cm}
\centerline{\includegraphics[width=0.8\textwidth]{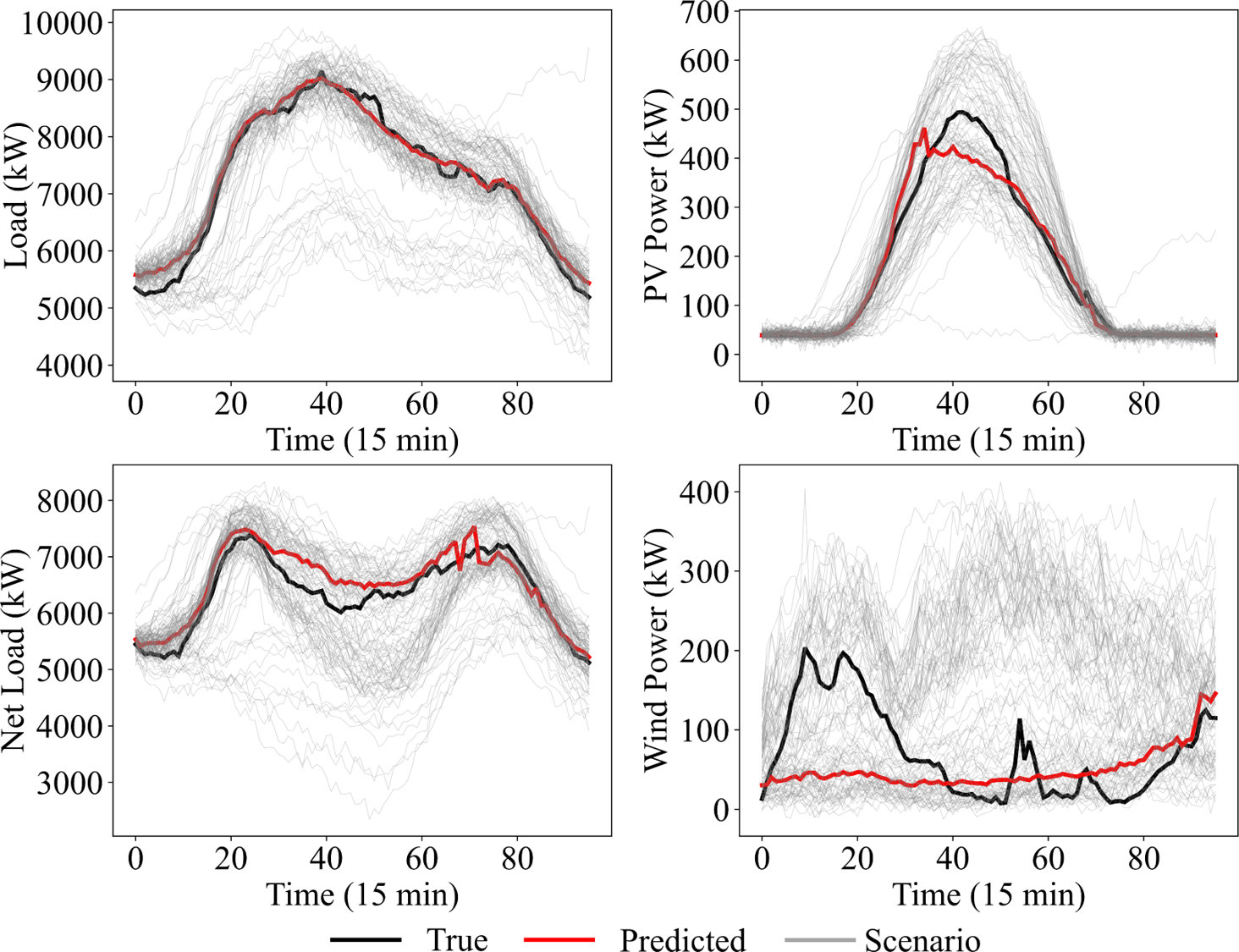}}
\caption{Day-ahead net load scenario generation results.}
\label{fig:multi_result}
\vspace{-0.2cm}
\end{figure}

\begin{figure*}[!h]
\vspace{-0.2cm}
\centerline{\includegraphics[width=1.0\textwidth]{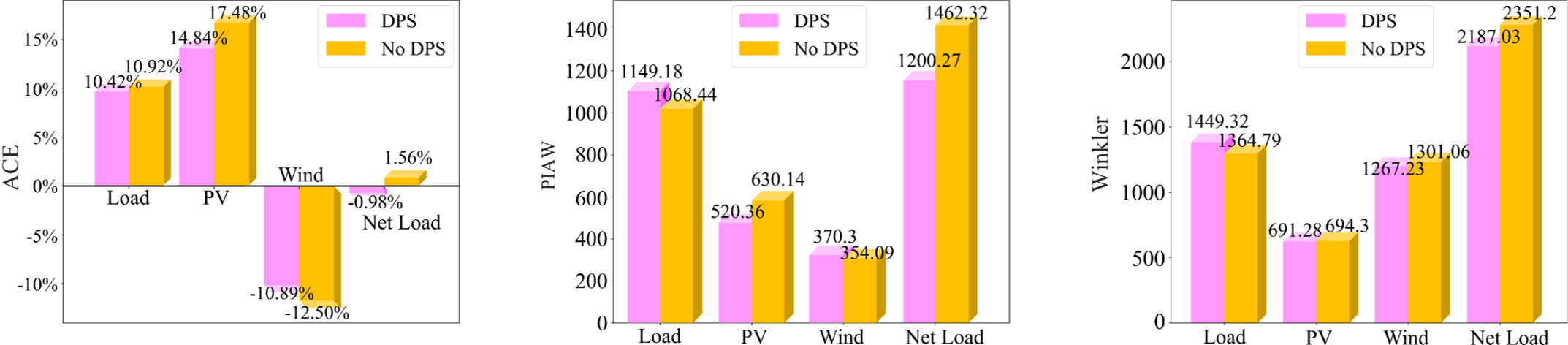}}
\caption{Probabilistic Forecast Results of Multi-Energy Forecast.}
\label{fig:multi}
\vspace{-0.4cm}
\end{figure*}

\subsection{The Effect of Multi-Energy Forecast}
This subsection presents a comprehensive analysis of multi-energy forecasting. Fig. \ref{fig:multi_result} illustrates the forecast outcome for multiple energy types, where the generative scenarios encompass the actual observations. Fig. \ref{fig:multi} showcases the results of employing the DPS method in the multi-energy forecast framework. The multi-energy forecast demonstrates superior probabilistic forecasting performance across net load, photovoltaic, wind power, and load forecasts, with ACE values approaching 0 and reduced Winkler scores. Notably, in net load forecasting, the ACE is -0.98\%. The variability introduced by RES is considerable, yet the DPS method effectively leverages the interrelationships among components to generate joint scenarios. The ACE for RES forecasts is larger due to their inherent stochastic nature, resulting in a broader interval for load forecasts. Despite the advantages seen with the DPS method over those without it, further improvements in the accuracy of multi-energy probabilistic forecasting are necessary, indicating a promising future research direction.

% \begin{table}\small
% \caption{Probabilistic Forecast Results of Multi-Energy Forecast}
% \label{tab_dps}
% \centering
% % \resizebox{1.0\linewidth}{!}{
% \begin{tabular}{ccccccccccc}
% \hline
% \multicolumn{2}{c}{} & \textbf{DPS} & {\textbf{No DPS}}\\
% \hline
% \multirowcell{2}{Load} & ACE  & 4.10\% & 1.98\%\\
%                           & PIAW & 0.7766 & 0.7618\\
% \hline
% \multirowcell{2}{Photovoltaic} & ACE  & 13.04\% & 11.56\% \\
%                           & PIAW & 0.6054 & 0.5564 \\
% \hline
% \multirowcell{2}{Wind Power} & ACE  & -0.1006\% & -0.0578\% \\
%                           & PIAW & 1.6883 & 1.3400 \\
% \hline
% \multirowcell{2}{Net Load} & ACE  & -1.78\% & -2.103\% \\
%                           & PIAW & 0.7925 & 0.7939 \\
% \hline
% \end{tabular}
% % }
% \end{table}

%section Case Studies (end)

\section{Conclusion} % (fold)
\label{sec:Conclusion}
The integration of distributed renewable energy sources (RES) increases the complexity of net load forecasting, requiring advanced techniques to manage uncertainty and volatility. The enhanced conditional diffusion model (ECDM) is proposed to address this by capturing forecast uncertainty and net load dynamics, using a conditional diffusion model with a cross-attention mechanism incorporating external conditions like weather and calendar data and an imputation-based sampling guided by historical net load. ECDM improves forecast accuracy through a weekly arrangement and enhances generative diversity by blending scenarios from unconditional models. Evaluated on a public dataset, ECDM demonstrates superior accuracy, reliability, sharpness, and uncertainty quantification compared to existing methods, with ablation experiments confirming the effectiveness of these enhancements. Future research will refine ECDM for more precise probabilistic multi-energy forecasts and explore novel denoising neural networks to boost its application in energy time series forecasting.

 \bibliographystyle{elsarticle-num} 
 \bibliography{references}




% %% For numbered reference style

%   \textit{\LaTeX: a document preparation system},
%   Addison Wesley, Massachusetts,

% \end{thebibliography}
\end{document}